\documentstyle[12pt,epsf]{article}

\begin{document}


\begin{center}

 {\Large \bf
\vskip 7cm
\mbox{Do Quark Mass Effects Survive}
\mbox{in the High-$Q^2$ Limit of DIS?}
}
\vskip 3cm

\mbox{A.V.~Kisselev, V.A.~Petrov, R.A.~Ryutin}

\mbox{{\small Institute for High Energy Physics, Division of Theoretical Physics,}}

\mbox{{\small{\it 142 281} Protvino, Russia}}

 \vskip 1.75cm
{\bf
\mbox{Abstract}}
  \vskip 0.3cm

\newlength{\qqq}
\settowidth{\qqq}{Quark mass effects are analyzed at high $Q^2$ in the current fragmentation region 
of DIS.}
\hfill
\noindent
\begin{minipage}{\qqq}
Quark mass effects are analyzed at high $Q^2$ in the current fragmentation region 
of DIS. It is found that the linear combination $F_2-2.75F^c_2$ scales at
large $Q^2$ and small $x$. We obtained a lower bound for the ratio $F^c_2/F_2$ 
which lies very close to the data from HERA.
\end{minipage}


\vskip 0.5cm
{\bf
\mbox{Keywords}}
\vskip 0.3cm

\settowidth{\qqq}{Quark mass effects are analyzed at high $Q^2$ in the current fragmentation region 
of DIS.}
\hfill
\noindent
\begin{minipage}{\qqq}
Heavy Quark Production -- Coefficient Functions -- Specific Scaling
\end{minipage}

\end{center}

\setcounter{page}{1}
\newpage


\section{Introduction}
\label{intro}

 The data on open-charm production in Deep Inelastic Scattering (DIS)
from HERA collider~\cite{experiments,experiments2} show that
its contribution, $F^c_2$, to the structure function 
runs up to $40\%$ at measured values of $x$ and $Q^2$ and increases faster than
$F_2$ when $x$ becomes smaller. The contribution of b-quarks, $F^b_2$, to the total 
structure function is of the order $2$-$3\%$, as one can see from recent measurements of 
open-beauty production~\cite{b_small}.

 One often believes that, with increasing energy of colliding particles, $W$, and 
momentum transfer squared $Q^2$, mass effects become 
insignificant. However, arguments were given in~\cite{Petrov1,Petrov2} 
that the difference between the DIS structure functions with open heavy quark production
in the current fragmentation region and structure functions of the process without heavy 
flavours scales for large $Q^2$, i.e. it depends on the Bjorken variable
$x$ and heavy quark mass $m_Q$ only. This result gives a possibility to obtain
model independent (i.e. independent on the concrete choice of the gluon distribution
inside the nucleon) lower bound for $F^c_2$, which is in an agreement with the
experimental data on $F^c_2$~\cite{Petrov1,Petrov2}.

In the present work we analyse the problem of mass scale influence on the behaviour of
physical quantities in the framework of the operator product expansion~\cite{OPE}.

 It is convenient to introduce the quantity
\begin{equation}
\label{eq1}
F^{Q\bar{Q}}_2=F^Q_2/e^2_Q
\end{equation}
instead of $F^Q_2$ ($Q=c,b$).
Analogously for the light quarks $q=u,d,s$ (that we treat massless)
we define
\begin{equation}
\label{eq2}
F^{q\bar{q}}_2=F^q_2/e^2_q\;{,}
\end{equation}
where $e_{Q(q)}$ are quark electric charges.

 The operator product expansion gives the following expression for the $F^{Q\bar{Q}}_2$:
\begin{eqnarray}
\label{int1}
\frac{1}{x}F^{Q\bar{Q}}_2(x,Q^2,m^2_Q)&=&C_g(\frac{Q^2}{\mu^2},\frac{m^2_Q}{\mu^2},\alpha_s(\mu^2))
\otimes f_g(\mu^2)[x] +\\
 &+& C_Q(\frac{Q^2}{\mu^2},\frac{m^2_Q}{\mu^2},\alpha_s(\mu^2))
\otimes f_Q(\mu^2)[x] +\nonumber\\
 &+& C_q(\frac{Q^2}{\mu^2},\frac{m^2_Q}{\mu^2},\alpha_s(\mu^2))
\otimes f_q(\mu^2)[x]\;{.}\nonumber
\end{eqnarray}
In Eq.~(\ref{int1}) the quantities $C_i$ are coefficient functions, $f_i$ are matrix elements of
corresponding composite operators, which can be identified with distributions 
of quarks and gluons inside the nucleon, and $\mu$ is a renormalization scale
of the composite operators.
Symbol $\otimes$ means convolution in the variable $x$,
\begin{equation}
\label{svertka}
a\otimes b[x]=\int\limits_x^1\frac{dy}{y}a(y)b\left(\frac{x}{y}\right)\;{.}
\end{equation}
In Eq.~(\ref{int1}) we have neglected higher twists contributions. To simplify notations, we 
do not show below $\alpha_s$-dependence of the coefficient functions. 

 It is well known that $C_i$ and $f_i$ separately depend on a renormalization 
scheme. Usually, the $\overline{{\rm MS}}$-scheme is more preferable than the MOM-scheme, because of
more complicated calculations in the framework of the latter.

 Nevertheless, the MOM-scheme has some advantages. One of them is the universality of the
calculational algorithm for coefficient functions in any order of perturbation 
theory. Advantages of use of the scheme with momentum subtraction in the case of
heavy quarks were noted in~\cite{old_scheme}, where 
authors proposed a mixed, so called, CWZ renormalization scheme.

The purpose of the present work is to show that effects related to the 
heavy quark mass $m_Q$, "survive" in the large $Q^2$ limit. There is a linear combination
of the structure functions $F_2$ and
$F^c_2$ that has scaling properties. It appears
that this scaling takes place in different renormalization schemes. The lower bound for the ratio 
$F^c_2/F_2$ as the function of $x$ at fixed values of $Q^2$ is also calculated. The results
are compared with the experimental data.

\section{Asymptotic relations between structure functions}
\label{sec:1}

 We are interested in the behaviour of $F^c_2$ at large $Q^2$ and
small $x$. We suppose that the production of heavy quarks in this region
results from gluons, and heavy quarks are not included in the evolution of
light quarks and gluons.

 So, we can write
\begin{eqnarray}
\label{c6a}
\frac{1}{x}F_2^{Q\bar{Q}}(x,Q^2,m^2_Q)&=&C_g(\frac{Q^2}{\mu^2},\frac{m^2_Q}{\mu^2})\otimes f_g(\mu^2) [x] +\\
&+&C_Q(\frac{Q^2}{\mu^2},\frac{m^2_Q}{\mu^2})\otimes f_Q(\mu^2) [x]\;{.}\nonumber
\end{eqnarray}
 Taking $\mu^2=\mu^2_0$, where $\Lambda^2_{\mbox{\small QCD}}\ll\mu^2_0\ll Q^2$, 
and neglecting "intrinsic charm" ("beauty") inside the nucleon, i.e.
assuming, that there is a scale $\mu^2_0$, at which heavy quark distribution function
is small as compared to the gluon one, we simplify Eq.~(\ref{c6a}) to the relation
\begin{equation}
\label{c6}
\frac{1}{x}F_2^{Q\bar{Q}}(x,Q^2,m^2_Q)=C_g(\frac{Q^2}{\mu^2_0},\frac{m^2_Q}{\mu^2_0})\otimes
f_g(\mu^2_0) [x]\;{.}
\end{equation}

Let us now define the quantity
\begin{equation}
\label{c7a}
\Delta F_2=F_2^{q\bar{q}}(x,Q^2)-F_2^{Q\bar{Q}}(x,Q^2,m^2_Q)\;{.}
\end{equation}
Assuming that at small $x$ the main contribution to the DIS structure function
without heavy quark production is determined by gluons, i.e. by the formula
analogous to~(\ref{c6}), we find
\begin{equation}
\label{c7}
\frac{1}{x}\Delta F_2=\Delta C_g\otimes f_g(\mu_0^2) [x]\;{,}
\end{equation}
where
\begin{equation}
\label{c8}
\Delta C_g=C_g(y, \frac{Q^2}{\mu_0^2}, 0)-
C_g(y, \frac{Q^2}{\mu^2_0}, \frac{m^2_Q}{\mu_0^2})\;{.}
\end{equation}

 Calculations of the gluon coefficient function in the order $O(\alpha_s)$ in the MOM-scheme
were given in~\cite{m_e_last}. Using the expression obtained there, we get from~(\ref{c8}):
\begin{eqnarray}
\label{deltaCg}
\Delta C_g \simeq \Delta C^{(1)}_g(y,\frac{m^2_Q}{\mu^2_0})=\frac{\alpha_s}{8\pi}\left\{
\left( y^2+(1-y)^2 \right)\ln\left[
1+\frac{m^2_Q}{\mu_0^2y(1-y)}
\right]-\right.\\
-\left.\frac{m^2_Q\left(1+2y(1-y)\right)}{m^2_Q+\mu_0^2y(1-y)}\right\}\;{.}\nonumber
\end{eqnarray}
Thus, the quantity $\Delta F_2$ tends to the finite 
limit $\Delta F_2(x,m^2_Q)$ at large $Q^2$. It was shown in~\cite{m_e_last}, that this result
is not the artifact of the MOM-scheme, although the very expression for the quantity
$\Delta '_g$ depends, of course, on the renormalization scheme.

 Starting from the expression for $\Delta C_g$, Eq.~(\ref{deltaCg}), it is easy to
see, that for $y \leq 0.1$ (small $y$ region is most important in the
analysis of the behavior of the structure function at $x\ll 1$)
\begin{equation}
\label{ineq1}
\Delta C_g>0\;{.}
\end{equation}
It follows 
from the explicit form of $C_g(y,\frac{Q^2}{\mu^2})$ in the MOM-scheme for 
the massless case in the order $O(\alpha_s)$ for any $Q^2$ and $\mu^2$ that~\cite{m_e_last} 
\begin{equation}
\label{c5a}
\left.C_g(y,\frac{Q^2}{\mu^2_0})\right|_{Q^2=m^2_Q}>\Delta C_g(y,\frac{m^2_Q}{\mu_0^2})\;{.}
\end{equation}
Then from Eqs.~(\ref{c7}), (\ref{ineq1}), (\ref{c5a}) we conclude:
\begin{equation}
\label{c5}
\left.F_2^{q\bar{q}}(x,Q^2)\right|_{Q^2=m^2_Q}>\Delta F_2(x,m^2_Q)>0\;{.}
\end{equation}

Inequalities~(\ref{c5}) are used below to obtain the lower bound for the ratio 
$F^c_2/F_2$.


\section{Charm contribution to the structure function}
\label{sec:2}

 It has been found in the previous section (see Eqs.~(\ref{c7}),(\ref{deltaCg})), that the difference
between contributions of light and heavy flavours to the DIS structure
function scales for
$Q^2\to\infty$.

 Taking this result as a basis, it is easy to see, that the following linear combination~\cite{Petrov2} 
\begin{eqnarray}
\label{c12}
\Sigma_{\alpha}(x,Q^2)&\equiv& F_2(x,Q^2)+\alpha F^c_2(x,Q^2,m^2_c)-\\
&-&(4\alpha +11)F^b_2(x,Q^2,m^2_b)\;\nonumber
\end{eqnarray}
has scaling behaviour, with
$\alpha$ being an arbitrary constant.

 To cancel the contribution of b-quarks from~(\ref{c12}), we have taken $\alpha=-2.75$. Then
we obtain the prediction, that the linear combination 
\begin{equation}
\label{sigma0}
\Sigma=F_2-2.75 F^c_2
\end{equation}
must tend to a function, that depends only on the Bjorken variable $x$ (and the heavy quark 
mass) for $Q^2\to\infty$.

 Using the expression for $\Delta C_g$  in the leading order in $\alpha_s$, we find, that the above difference~(\ref{sigma0})
tends to the scaling limit in the region $m^2_Q\ll Q^2$ in the following way 
\begin{eqnarray}
\label{c14}
\frac{1}{x}\Sigma &=& \frac{1}{9}\left[
7\Delta C^{(1)}_g(\frac{m^2_c}{\mu_0^2})-\Delta C^{(1)}_g(\frac{m^2_b}{\mu_0^2})
\right]\otimes f_g(\mu_0^2) [x]+\\
 &+&\frac{m^2_b-7m^2_c}{Q^2}\ln(\frac{Q^2}{\mu^2_0})\cdot h\otimes f_g(\mu_0^2) [x]\;{,}\nonumber
\end{eqnarray}
where
\begin{equation}
\label{c15}
h(y)=\frac{1}{9}y(1-y)[(2-3y)^2 + 3y^2]\;{.}
\end{equation}
Since $m^2_b-7m^2_c>0$, we conclude, that the correction in the expression for
$\Sigma (x,Q^2)$, (\ref{c14}), is positive and decreasing on $Q^2$.

 To compare our results with the recent experimental data, we paramet\-ri\-ze 
$F^c_2$ according to the expression~(\ref{c14}) for $\Sigma(x,Q^2,m^2_Q)$ and fit the data
from HERA collider~\cite{experiments2}. 

 For $F_2$ we use the parametrization of H1 collaboration~\cite{experiments}:
\begin{equation}
\label{F2}
\!\!\!\!\!\; F_2(x,Q^2)=\left[ a x^b + c x^d (1 + e\sqrt{x})\left(\ln Q^2 + f\ln^2Q^2 + \frac{h}{Q^2}
\right)\right](1-x)^g\;{,}
\end{equation}
with parameters
\begin{center}
\begin{tabular}{|c|c|c|c|}
\hline
a & b & c & d\\
\hline
3.1 & 0.76 & 0.124 & $-$0.188\\
\hline
\hline
e & f & g & h\\
\hline
$-$2.91 & $-$0.043 & 3.69 & 1.4 GeV$^2$\\
\hline
\end{tabular}
\end{center}

 For $F^c_2$ we choose the expression qualitatively coherent to the asymptotic
behaviour of the quantity $\Sigma =F_2 - 2.75 F^c_2$ in the variable
$Q^2$~(\ref{c14})
\begin{equation}
\label{F2c}
F^c_2(x,Q^2)=\frac{1}{2.75}F_2(x,Q^2) -\bar{a}x^{\bar{b}}(1-x)^{\bar{g}}
\left[
1+x^{\bar{c}}\frac{\bar{h}}{Q^2}\ln Q^2
\right]\;{,}
\end{equation}
where $F_2$ is defined above~(\ref{F2}). Fitting the data from HERA~\cite{experiments2} 
for $6.5$~GeV$^2\le Q^2\le 130$~GeV$^2$ gives
the following values of parameters
\begin{center}
\begin{tabular}{|c|c|c|c|c|}
\hline
$\bar{a}$ & $\bar{b}$ & $\bar{c}$ & $\bar{g}$ & $\bar{h}$ \\
\hline
0.28 & 0.15 & $-$0.08 & 5.00 & 1.86 GeV$^2$ \\
\hline
\end{tabular}
\end{center}
and $\chi^2/n.d.f.=34.6/36=0.96$.

 In Fig. 1 we show the dependence of the quantity $\Sigma$ as a function of $Q^2$
for two values of $x$, for which we have the array of experimental data, obtained at
different $Q^2$ and $x$, closest to the given values 
$x=0.01$ and $x=0.001$. As we see, experimental data are in a good agreement with our result
on the approach to scaling behaviour from above but the very existence of the scaling may be checked
at higher $Q^2$.

 As has been found in~\cite{Petrov2}, inequalities~(\ref{c5}) allow us to obtain
the following estimation for the ratio of structure functions:
\begin{equation}
\label{c16}
\frac{F^c_2(x,Q^2)}{F_2(x,Q^2)}>0.4\left( 1-\frac{F_2(x,m^2_c)}{F_2(x,Q^2)}\right)\;{.}
\end{equation}
It is important to stress, that we did not use any parametrization for $F^c_2$ to obtain
the inequality, and also, that it does not depend on the behaviour of the gluon
distribution. 

 Curves represented in Figs. 2, 3 are calculated by using the 
formula~(\ref{c16}) for two values of
$c$-quark mass and are compared with the data of ZEUS 
collaboration~\cite{experiments2}. Despite the fact that these curves are only lower bounds
for the ratio $F^c_2/F_2$, they lay very close to the experimental points.

 Our estimations show that lower-bound curves are also in a good agreement
with new preliminary data of ZEUS collaboration~\cite{experiments3}, including the data for
maximal measured value $Q^2=565$~GeV$^2$.


\section{Conclusions}
\label{concl}

 In the present work we analyzed quark mass
effects in DIS with the help of OPE. It is shown by calculating in the leading order 
that the new scaling takes place in DIS; certain linear combination
of the DIS structure function and the DIS structure function with open charm production
scales in the limit of large $Q^2$.

 It is found that this specific scaling is in a good agreement with the $Q^2$-trend of recent
experimental data on $F^c_2$ and $F_2$ obtained at HERA. 

Certainly, higher orders
could change the theoretial conclusion about such a scaling. At present the
first order result can serve as an indication of the existence of an interesting 
physical phenomenon. Influence of higher orders is a subject of our further work. 

 We also calculated the lower bound for the ratio $F^c_2/F_2$ as a function of 
$x$ at fixed $Q^2$, independent on the shape of the gluon distribution in the nucleon, and
compared it with the data from ZEUS collaboration. This lower bound appears to be quite close
to the data.

 We are grateful to L.K.Gladilin for providing us with recent results of ZEUS collaboration on the ratio
$F^c_2/F_2$.




\newpage
\section*{Figure captions}

\begin{list}{Fig.}{}

\item 1: $Q^2$-dependence of the difference of structure functions at two
         fixed values of $x$. Solid curves are obtained by fitting the
		 data on $F^c_2$ from~\cite{experiments2,experiments3}. The experimental points
		 have $x$ closest to the given values $x=0.01$ and $x=0.001$.

\item 2: The ratio $F^c_2/F_2$ as the function of the variable $x$ at fixed values of
         $Q^2$. Dashed curves represent the lower bound for
         $F^c_2/F_2$ for the c-quark mass
         $m_c=1.7$~GeV, dotted curves correspond to $m_c=1.3$~GeV. Experimental 
		 points are taken from~\cite{experiments2}.

\item 3: The same as in Fig. 2, but for other values of $Q^2$.

\end{list}


\newpage

\begin{figure}[hb]
\label{fig1}
\vskip 2cm
\hskip  1cm \vbox to 19cm {\hbox to 16cm{\epsfxsize=16cm\epsfysize=19cm\epsffile{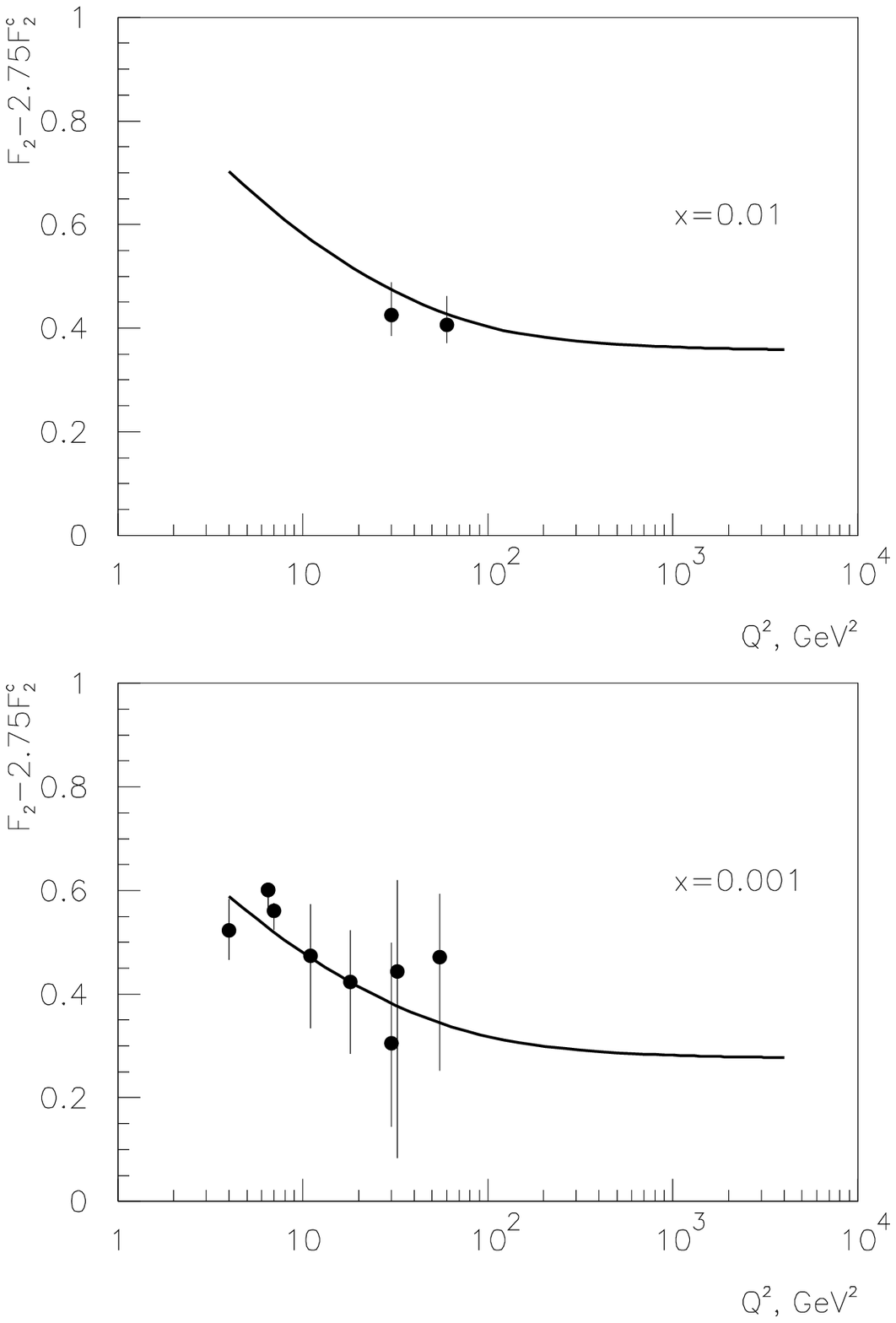}}}
\hskip 1cm
\caption{}
\end{figure}

\newpage

\begin{figure}[hb]
\label{fig2}
\vskip 1.5cm
\vbox to 17cm {\hbox to 17cm{\epsfxsize=17cm\epsfysize=17cm\epsffile{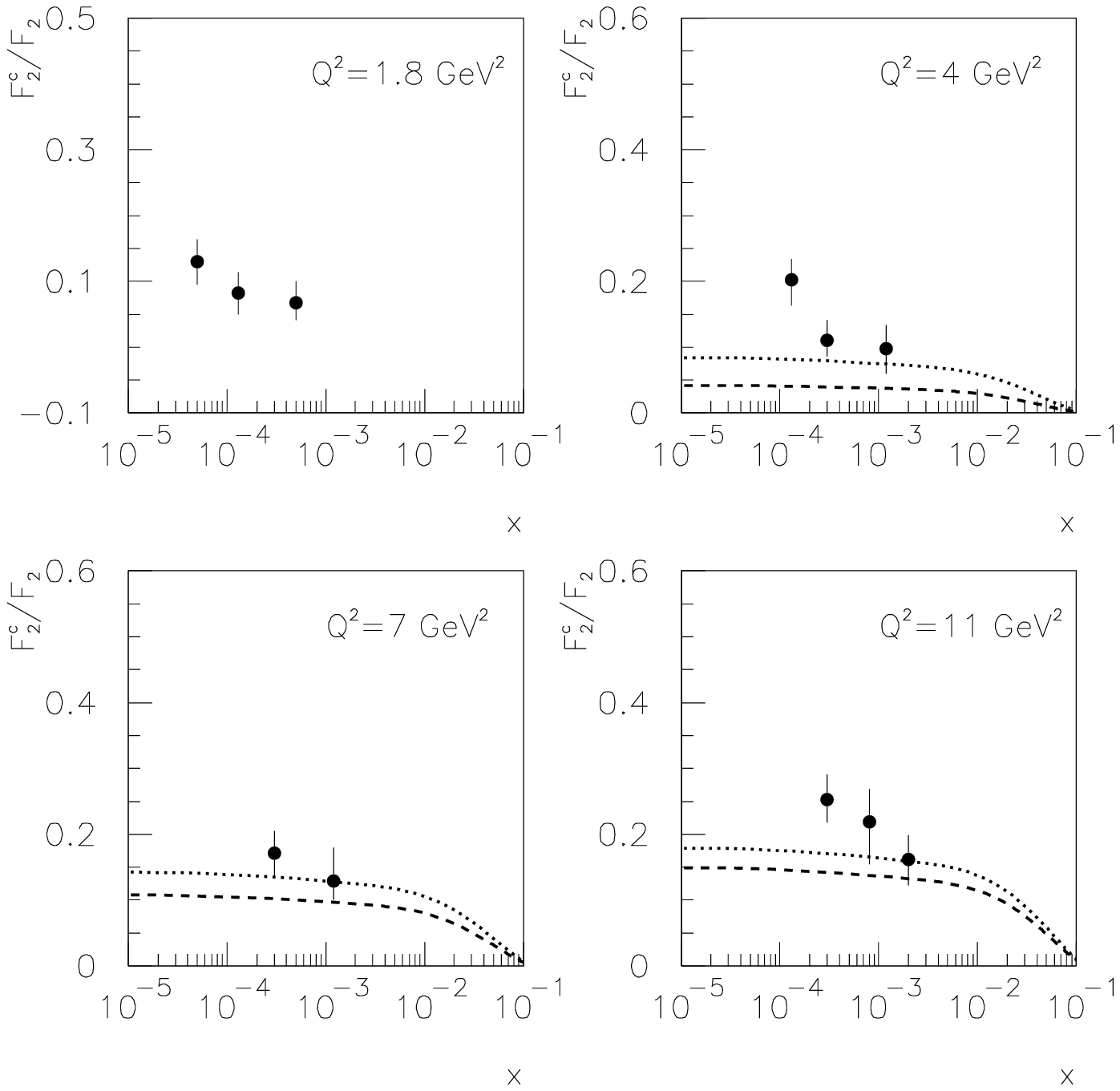}}}
\hskip 1cm
\caption{}
\end{figure}

\newpage

\begin{figure}[hb]
\label{fig3}
\vskip 1.5cm
\vbox to 17cm {\hbox to 17cm{\epsfxsize=17cm\epsfysize=17cm\epsffile{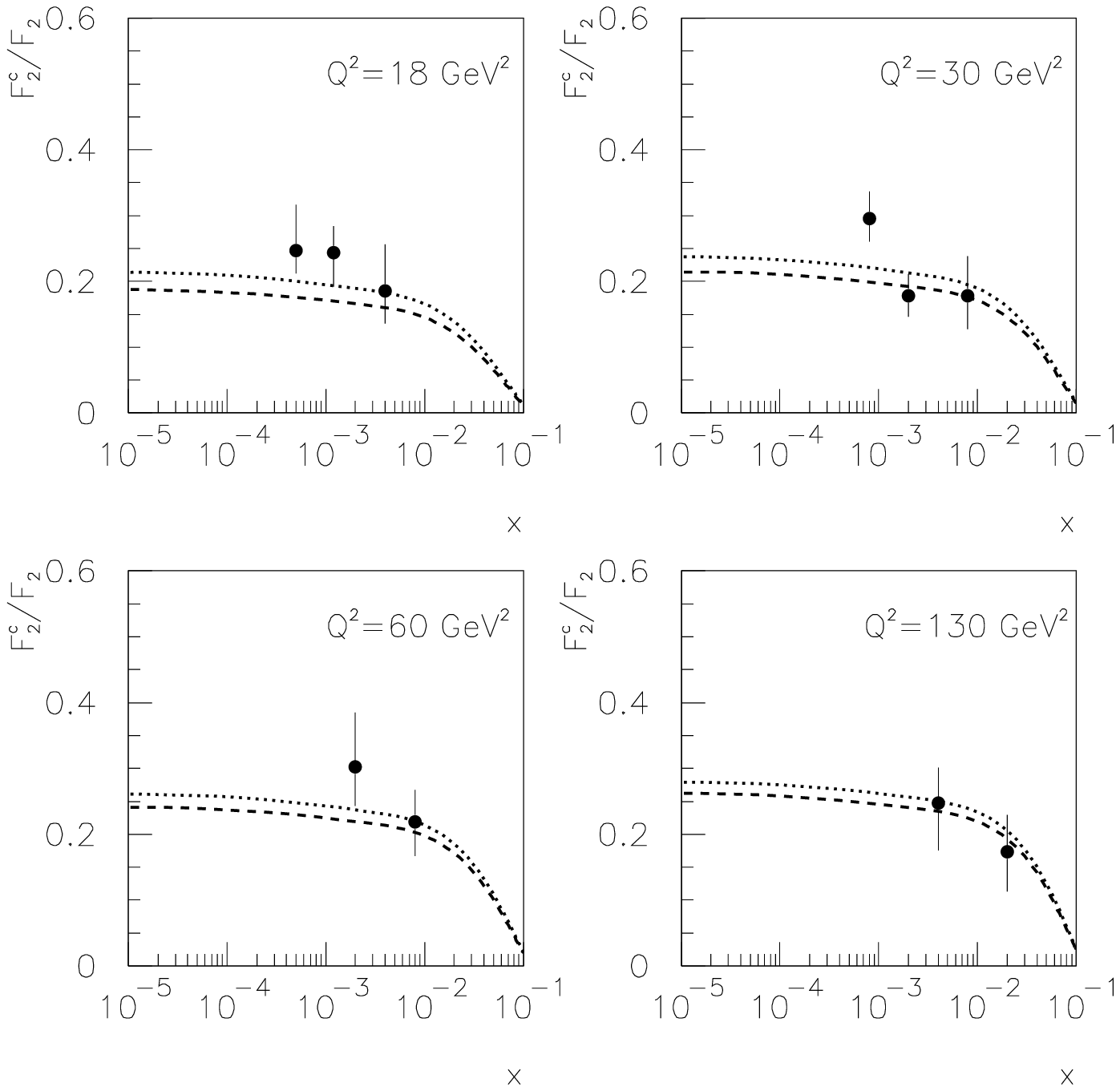}}}
\hskip 1cm
\caption{}
\end{figure}

\end{document}